\documentclass[nohyper,notoc]{JHEP}

\usepackage{amsmath,euscript,array,amssymb,cite} 
\usepackage{mathrsfs}

\def\N{{\cal N}}
\def\cN{{\cal N}}

\def\Tr{{\rm Tr}}

\def\Dbarslash{\,\,{\raise.15ex\hbox{/}\mkern-12mu {\bar\D}}}
\def\Dslash{\,\,{\raise.15ex\hbox{/}\mkern-12mu \D}}
\def\delslash{\,\,{\raise.15ex\hbox{/}\mkern-9mu \partial}}
\def\delbarslash{\,\,{\raise.15ex\hbox{/}\mkern-9mu {\bar\partial}}}

\def\b{\beta}         
      
\def\d{\delta}  \def\D{\Delta}

\def\l{\lambda} \def\L{\Lambda}

\def\s{\sigma}    
\def\t{\tau}  
\def\th{\theta}

 \def\cB{{\cal B}}

\def\cM{{\cal M}} 
\def\cN{{\cal N}}   
\def\cP{{\mathscr P}}

\newcommand{\eq}[1]{(\ref{#1})}

\def\bx{{\bf x}}  
\def\CC{\mathbb{C}}

\def\l{\lambda} \def\L{\Lambda}
\def\t{\tau}
\def\s{\sigma}

\def\D{{\cal D}}
\def\Dbarslash{\,\,{\raise.15ex\hbox{/}\mkern-12mu {\bar\D}}}
\def\delslash{\,\,{\raise.15ex\hbox{/}\mkern-9mu \partial}}
\def\Dslash{\,\,{\raise.15ex\hbox{/}\mkern-12mu \D}}

\def\adss{AdS_5\times S^5}
\def\ads{AdS_5}

\newcommand{\AdS}{AdS}

\newcommand{\be}{\begin{equation}}
\newcommand{\ee}{\end{equation}}
\def\bea{\begin{eqnarray}}
\def\eea{\end{eqnarray}}

\def\tb{\bar{\tau}}
\def\bb{\bar{\beta}}

\title{String theory dual of the $\beta$-deformed gauge theory}

\author{Chong-Sun Chu$^a$ and Valentin V.~Khoze$^b$\\
$^a$Department of Mathematical Sciences, University of Durham,
Durham, DH1 3LE, UK\\
$^b$Department of Physics and IPPP, University of Durham,
Durham, DH1 3LE, UK\\
E-mails: {\tt chong-sun.chu@durham.ac.uk,} $\,$
{\tt valya.khoze@durham.ac.uk}}

\abstract{We consider the AdS/CFT correspondence between the
$\beta$-deformed supersymmetric gauge theory and the type IIB string
theory on the Lunin-Maldacena background. Guided by gauge theory
results, we modify and extend the supergravity solution of Lunin and
Maldacena in two ways. First we make it to be doubly periodic in the
deformation parameter, $\beta \to \beta+1$ and $\beta \to \beta+\tau_0,$
to match the $\beta$-periodicity property of the dual gauge theory.
Secondly, we reconcile the $SL(2,Z)$ symmetry of the gauge theory, which
acts on the constant parameters $\tau_0$ and $\beta$, with the $SL(2,Z)$
invariance of the string theory, which involves the dilaton-axion field
$\tau (\bx)$. 
Our proposed modified configuration transforms correctly under the
$SL(2,Z)$ of string theory when its parameters are transformed under the
$SL(2,Z)$ of the gauge theory. We interpret the resulting configuration
as the string theory (rather than supergravity) background which is dual
to the $\beta$-deformed conformal Yang-Mills. Finally, we check that our
string theory background leads to the IIB effective action which is
correctly reproduced by instanton calculations on the gauge theory side,
carried out at weak coupling, in the large-$N$ limit, but to all orders
in the deformation parameter $\beta$.}

\preprint{{\tt hep-th/0603207}
\\IPPP/06/19\\
DCPT/06/38
}

\begin{document}

%%%%%%%%%%%%%%%%%%%%%%%%%%%%%%%%%%%%%%%%%%%%%%%%%%%%%%%%%%%%%%%%%%%%%%%%%%%%%%
\section{Introduction} 

The purpose of this paper is to discuss certain general properties of
the AdS/CFT correspondence between the exactly marginal $\beta$-deformed
gauge theories and the type IIB string theory. In its original
formulation, the AdS/CFT duality \cite{Maldacena} relates the string
theory on a curved background $AdS_5 \times S^5$ to the ${\cal{N}}=4$
supersymmetric Yang-Mills theory living on the boundary of $AdS_5$. All
symmetries are known to match precisely on the left and on the right
hand side of the correspondence. In particular, the $SL(2,Z)$ S-duality
of the ${\cal{N}}=4$ SYM becomes the $SL(2,Z)$ duality of IIB string
theory.

$\beta$-deformations of the $\cN=4$ supersymmetric Yang-Mills define a
family of conformally-invariant four-dimensional $\cN=1$ supersymmetric
gauge theories. The AdS/CFT duality extends to the $\beta$-deformed
theories where it relates the $\beta$-deformed $\cN=4$ SYM and the
supergravity on the deformed $\AdS_5\times \tilde{S}^5$ background. The
gravity dual was found by Lunin and Maldacena in Ref.~\cite{LM}, and
this provides a foundation for a precise formulation of the AdS/CFT
duality in the deformed case. In a recent paper \cite{GK} the
supergravity dual and the resulting string theory effective action were
successfully tested and studied using Yang-Mills instanton methods
developed earlier for the $\cN=4$ case in \cite{BGKR,DKMV,MO3,Review}. 
The approach of \cite{GK} was based
on the semiclassical (i.e. weak coupling) approximation  
and, in order to compare with the Lunin-Maldacena solution,
the results were further
restricted to the small-deformation limit $\beta \ll 1$ with $\beta^2 N
~$ fixed, in the usual large-$N$ limit.

It is known that the $\beta$-deformed gauge theory retains the $SL(2,Z)$
duality of the parent $\N=4$ SYM \cite{DHK}. At the same time, the IIB
string theory is invariant under the $SL(2,Z)$ transformations of string
theory. In the present paper we want to study the AdS/CFT correspondence
between the gauge and the string theory for general values of the
deformation parameter $\beta$, as well as the interplay between the
$SL(2,Z)$ duality on the gauge theory and on the string theory side.

Marginal deformations of the SYM have also been studied extensively in 
\cite{LS,BL,BJL,Aharony1,Aharony2}.
More recently,  
several perturbative calculations in the $\beta$-deformed theories were
carried out in \cite{FG,MPSZ,VVK} where it was noted that
there are many similarities between the deformed and the undeformed
theories which emerge in perturbation theory in the large number of colours
limit. In \cite{VVK} it was shown that for real values of $\beta$ all
perturbative scattering amplitudes in the $\beta$-deformed theory are
completely determined by the corresponding $\cN=4$ amplitudes.
Studies of perturbative integrability of $\beta$-deformations were
conducted in Refs.~\cite{integr1,integr2,integr3}.

%%%%%%%%%%%%%%%%%%%%%%%%%%%%%%%%%%%%%%%%%%%%%%%%%%%%%%%%%%%%%%%%%%%%%%%%%%%%%%%%%%%%%%%%%%%%%%%%%%%%%
%\subsection{Gauge theory formulation}

The $\beta$-deformed SYM is a conformal $\N=1$ supersymmetric gauge
theory obtained by an exactly marginal deformation of the superpotential
of the $\N=4$ SYM. In terms of the three adjoint chiral superfields of
the $\N=4$ theory, the deformation takes the form:
\be
\label{superpot2}
{\cal W}=\, i g \,\Tr( \Phi_1 \Phi_2 \Phi_3 -\Phi_1 \Phi_3 \Phi_2 )\, \to\,
i h \,\Tr( e^{ i \pi \beta } \Phi_1 \Phi_2 \Phi_3 - e^{-i \pi \beta } 
\Phi_1 \Phi_3 \Phi_2 )\ .
\ee
The deformed superpotential preserves one $\N=1$ supersymmetry 
of the original $\N=4$ SYM
and leads to a theory with a global $U(1)\times U(1)$ symmetry \cite{LM} 
\bea \nonumber
U(1)_1:& ~~~~&(\Phi_1,\Phi_2 ,\Phi_3 ) \to 
(\Phi_1,e^{i\varphi_1}\Phi_2 ,e^{-i \varphi_1} \Phi_3 )\ ,
\\ \label{u1s}
U(1)_2:&~~~~&(\Phi_1,\Phi_2 ,\Phi_3 ) \to 
(e^{-i \varphi_2} \Phi_1,e^{i\varphi_2}\Phi_2 , \Phi_3 )\ .
\eea
At the classical level the deformation in \eqref{superpot2} is marginal 
(the superpotential has classical mass dimension three)
and the deformed theory is parameterized by three complex constants,
$h, \beta, \tau_0,$
with $\tau_0$ being the usual complexified gauge coupling.

At quantum level, this deformation is not exactly marginal since the
operators in \eqref{superpot2} can develop anomalous dimensions. Using
constraints of $\N=1$ supersymmetry, and the exact NSVZ beta function,
Leigh and Strassler \cite{LS} argued that the deformation
\eqref{superpot2} is marginal at quantum level subject to a {\it single}
complex constraint on the three parameters, $\gamma(h, \beta,
\tau_0)=0.$ Here the function $\gamma$ is the sum of the anomalous
dimensions $\gamma_i$ of the three fields $\Phi_i$, so that $\gamma=
\sum_{i=1,2,3} \gamma_i.$ This constraint implies that there is a
2-complex-dimensional surface $\gamma(h, \beta, \tau_0)=0$ of
conformally invariant $\N=1$ theories\footnote{For a special case when
the deformation parameter $\beta$ is real, and in the large $N$ limit,
the Leigh-Strassler constraint can be solved to all orders in
perturbation theory \cite{MPSZ,VVK} and gives simply $|h|^2=g^2.$ In
this paper we will retain general complex deformations for which no
simple solution of the constraint is known to all orders.} obtained by
deforming $\N=4$ SYM. From now on, we will always assume that the
Leigh-Strassler conformal constraint is formally resolved in terms of
$h=h(\tau_0, \beta)$, and that the parameter $h$ is eliminated. This
implies that the $\beta$-deformed theory is a conformal $\N=1$
supersymmetric gauge theory which is characterized by two mutually
independent complex constants, $\beta$ and $\tau_0$. 

Furthermore, it has been argued by Dorey, Hollowood and Kumar in
\cite{DHK} that the $\beta$-deformed SYM retains the $SL(2,Z)$ duality
of the parent $\N=4$ theory. More precisely, the authors of \cite{DHK}
showed that at least in the massive phase the deformed theory has an
action of the $SL(2,Z)$ group which interchanges various massive vacua
of the theory. Under this transformation, the deformation parameter
$\beta$ transforms as a modular form \cite{DHK}:
\be
\label{SYMsl2z}
\tau_0 \, \to\, \tau_0'\, =\, \frac{a\tau_0 +b}{c\tau_0+d} \ , \qquad 
\beta \, \to\, \beta'\,=\,\frac{\beta}{c\tau_0+d}\ , \qquad ad-bc=1,
\ a,b,c,d\in Z \ .
\ee
For the gauge coupling
$\tau_0$ to transform in the standard fractional-linear way 
\eqref{SYMsl2z}, it is required \cite{DHK} to define it in a particular
way, 
\be
\tau_0 \, =\,  {\theta \over 2\pi} \, +\, {4\pi i \over g^2} \,
-\,n\frac{iN}{\pi}\, \log\frac{h}{g} \ . 
\label{tau0coup} 
\ee
 which we will always assume.\footnote{This parameter $h$ in the shift
 on the right hand side of  \eqref{tau0coup} is precisely the  
 coefficient in front of the superpotential in the deformed theory.
 In the undeformed theory this additional shift disappears since $h=g$.}

The action of the $SL(2,Z)$ transformations in the deformed gauge theory
must then persists also in the conformal phase. In what follows, we will
call the action of $SL(2,Z)$ in Eq.~\eqref{SYMsl2z} -- the SYM $SL(2,Z)$
duality. The relation of this duality with the $SL(2,Z)$ duality of the
corresponding type IIB string theory turns out to be non-trivial for
$\beta \neq 0$. The relation and reconciliation between the two
$SL(2,Z)$ transformations on the SYM and on the string side is one of
the main subjects of this paper.

The $\beta$-deformed gauge theory is also known to be doubly
periodic in $\beta$. First, it is obvious from the deformation of the
superpotential in \eqref{superpot2} that all observables 
in the theory must be periodic as
$\beta \to \beta+1$. In fact, all known perturbative and
multi-instanton calculations \cite{GK} in this theory automatically exhibit this
periodicity. Secondly, it was argued in \cite{DHK} that the gauge theory
is also periodic under $\beta \to \beta + \tau_0$. 
Clearly this double periodicity in $\beta$ must also be manifest in
the dual string theory formulation. 
However, the $\beta$-periodicity in string theory is obscured in
the supergravity description. One must use the full string theory
background rather than the supergravity solution, which 
is a good approximation to string theory only for small values of
$\beta$  \cite{LM}. In particular,  
the supergravity
dual of Lunin-Maldacena does not exhibit the double $\beta$ periodicity. 
The reconciliation of the $\beta$-periodicity 
on the SYM and on the string side is the second motivation of this paper.

%%%%%%%%%%%%%%%%%%%%%%%%%%%%%%%%%%%%%%%%%%%%%%%%%%%%%%%%%%%%%%%%%%%%%%%%%%%%
\section{Transformation properties of the gravity dual} 

The supergravity AdS/CFT dual of the $\beta$-deformed gauge theory was
constructed by Lunin and Maldacena \cite{LM} by applying a solution generating
$SL(3,R)$ transformation to the $\adss$ background, or equivalently by
a STsTS${}^{-1}$
transformation.
As its gauge theory dual, the supergravity solution depends
on the two complex parameters, $\tau_0$ and $\beta$, which give four
real constants. In describing the Lunin-Maldacena solution as well as
its string theory generalization we will denote real and imaginary parts
of these parameters as 
\be
\label{t0b12}
\tau_0 \, =\, \tau_{01}\, +\, i \tau_{02}\, , \qquad
\beta \, =\, \beta_{1}\, +\, i \beta_{2}\ .
\ee
The fifth real parameter of the supergravity solution is
the (quantized) radius $R_E$ 
which in $\sqrt{\alpha'}$ units is
\be
R_E^4 \, =\, 4\pi N \, \gg\, 1\ .
\ee
We further define functions $Q$, $g_{0E}$, $G$ and $H$ which depend 
on the coordinates
$\mu_i$ of the deformed sphere $\tilde{S}^5$,
\bea
\label{Qdef}
Q\, := \, \mu_1^2 \mu_2^2 + \mu_2^2 \mu_3^2 +\mu_1^2 \mu_3^2\ , \qquad
g_{0E}\, := \, R_E^4 \, Q \ , \\
\label{GHdef}
G^{-1} \, := \,1+ \frac{|\beta|^2}{\tau_{02}} g_{0E} \ , \qquad \quad
H\, := \,1+ \frac{{\beta_2}^2}{\tau_{02}} g_{0E}\ . 
\eea
The Lunin-Maldacena solution, Eqs.~(3.24)-(3.29) of Ref~\cite{LM},
is written in terms of these functions.
The metric in the Einstein frame is the warped product of 
the $\ads$ factor and the deformed 5-sphere, $\tilde{S}^5$,
\be
\label{metrdef}
ds^2_{E} \,=\, R_E^2\, G^{-1/4} \left[ ds^2_{AdS_5} +  
\sum_{i=1}^3 ( d\mu_i^2  + G \mu_i^2 d\varphi_i^2) 
+ \frac{|\beta|^2}{\tau_{02}}\,R_E^4\,G \mu_1^2\mu_2^2\mu_3^2 
(\sum_{i=1}^3 d\varphi_i)^2 \right] \ ,
\ee
where the sphere is parameterized by the three angles $\varphi_i$ and
the three radial variables $\mu_i$, 
which satisfy the condition $\sum_{i=1}^3\mu_i^2=1$. 
When $\beta=0$, the sphere is the undeformed $S^5.$

The dilaton, $\phi$, and the axion, $\chi$, fields of the Lunin-Maldacena solution 
are given by \cite{LM}
\be
\label{dilaxdef}
e^{-\phi} \, =\, \tau_{02}\, G^{-1/2}\, H^{-1} \ , \qquad
\chi\, =\, \tau_{01} \, -\beta_1\beta_2\,H^{-1}\,g_{0E}\ .
\ee
In addition, the  solution contains the two-form fields
(which were absent in the undeformed $\beta=0$ case)
\bea
\label{B2def}
B^{\rm NS}_2 \, &=&\,
\frac{\beta_1}{\tau_{02}}\,R_E^4\,G\,w_2 \, +\,12\,
\frac{\beta_2}{\tau_{02}}\,R_E^4\,w_1\,d\psi \ ,
\\
\label{C2def}
C_2 \, &=&\,
\left(\beta_2\,+\,\frac{\tau_{01}}{\tau_{02}}\,\beta_1\right) 
R_E^4\,G\,w_2\, -\, 
12\,
\left(\beta_1\,-\,\frac{\tau_{01}}{\tau_{02}}\,\beta_2\right)
\,R_E^4\,w_1\,d\psi \ ,
\eea
as well as the usual five-form field-strength
\be
\label{f5def}
F_5 \, =\, 4 R_E^4 \left(\omega_{\ads} \,+\, G\,\omega_{S^5}\right)\ .
% \ , \qquad\omega_{S^5} \, :=\, dw_1 d\varphi_1 d\varphi_2 d\varphi_3
\ee
The forms $w_1$, $w_2$, $d\psi$ and $\omega_{S^5}$ used in 
Eqs.~\eqref{B2def}-\eqref{f5def} above
involve only the coordinates on the deformed
sphere which can be found in \cite{LM}. These expressions will not 
be needed for our purposes.

It is important to stress that the dilaton and axion fields in \eqref{dilaxdef}
are not constant and should be distinguished from
the corresponding field theory constant values $\tau_{02}$ and $\tau_{01}$. 
This implies that the 
$SL(2,Z)$ symmetry of the IIB string theory, which acts on the dilaton-axion field 
$\tau$
\be
\label{strsl2z}
\tau \, \to\, \tau'\,=\, \frac{a\tau +b}{c\tau+d} \, , 
\qquad ad-bc=1,\ a,b,c,d\in Z \ , \qquad
\tau\,:=\, \chi +ie^{-\phi}\ .
\ee
should be distinguished from the $SL(2,Z)$ transformations on the gauge
theory side in \eqref{SYMsl2z} which act on the parameters in
\eqref{t0b12}. This distinction is specific to the $\beta$-deformed
theory, since for $\beta=0$ the dilaton-axion field $\tau$ was equal to
the SYM coupling $\tau_0$, and the two $SL(2,Z)$ symmetries were one and
the same.

\subsection{S-duality group}

We can now test how the Lunin-Maldacena solution above transforms under
the SYM $SL(2,Z)$ transformation of \eqref{SYMsl2z}. One would hope that
the solution would transform covariantly, and that the action of the
$SL(2,Z)$ of \eqref{SYMsl2z} on the parameters of the solution, would
give the original solution transformed under the string theory $SL(2,Z)$
of \eqref{strsl2z}. This would imply that the metric and the 5-form are
invariant, that the two 2-forms transform as a doublet, and that the
$\tau$-field transforms as in \eqref{strsl2z}. We will show now, that
this expectation holds on the Lunin-Maldacena solution {\it except} for
the $\tau$-field.

We first note that under the action of the SYM $SL(2,Z)$ of \eqref{SYMsl2z}, 
the functions $Q$, $g_{0E}$ and $G$ (but not $H$) in 
\eqref{Qdef}-\eqref{GHdef} are invariant:
\be
\label{QGinv}
Q(\tau_0',\beta')\, = \, Q(\tau_0,\beta)\ , \qquad
g_{0E}(\tau_0',\beta')\, = \, g_{0E}(\tau_0,\beta)\ , \qquad
G(\tau_0',\beta')\, = \, G(\tau_0,\beta)\ .
\ee
This implies that the metric \eqref{metrdef}
and the 5-form \eqref{f5def} of the Lunin-Maldacena solution are
also invariant under this $SL(2,Z)$, as expected,
\be
\label{metf5inv}
ds^2_{E} (\tau_0',\beta') \,=\, ds^2_{E}(\tau_0,\beta)\ , \qquad
F_5(\tau_0',\beta') \,=\, F_5(\tau_0,\beta)\ .
\ee
Furthermore, it can be shown that the two 2-forms \eqref{B2def}-\eqref{C2def}
do transform as a doublet under the SYM $SL(2,Z)$ of \eqref{SYMsl2z},
again in agreement with what is expected from the action of the string 
theory  $SL(2,Z)$:
\bea
\label{CBdoublet}
\begin{pmatrix}
C_2(\tau_0',\beta') \\ B^{\rm NS}_2(\tau_0',\beta')
\end{pmatrix} 
\, &=&\, 
\begin{pmatrix}
&a & b& \\ &c & d&
\end{pmatrix}
\,
\begin{pmatrix}
C_2(\tau_0,\beta) \\ B^{\rm NS}_2(\tau_0,\beta)
\end{pmatrix} \ .
\eea
To verify this transformation, it is convenient to use the fact that
the following combination of the parameters transforms as a doublet 
under \eqref{SYMsl2z},
\bea
\begin{pmatrix}
\beta_2 + \tau_{01} \beta_1 /\tau_{02}
 \\ \beta_1/\tau_{02}
\end{pmatrix} 
\, &\to& \, 
\begin{pmatrix}
&a & b& \\ &c & d&
\end{pmatrix}
\,
\begin{pmatrix}
\beta_2+\tau_{01}\beta_1/\tau_{02}
 \\ \beta_1/\tau_{02}
\end{pmatrix}  \ ,
\\
\nonumber
\\
\begin{pmatrix}
\beta_2/\tau_{02} \\
\beta_1-\tau_{01}\beta_2/\tau_{02}
\end{pmatrix} 
\, &\to& \, 
\begin{pmatrix}
&a & b& \\ &c & d&
\end{pmatrix}
\,
\begin{pmatrix}
\beta_2/\tau_{02} \\
\beta_1-\tau_{01}\beta_2/\tau_{02}
\end{pmatrix} \ .
\eea

It is, however, impossible to make the dilaton-axion field $\tau= \chi
+ie^{-\phi}$ in \eqref{dilaxdef} transform according to
Eq.~\eqref{strsl2z} by acting with \eqref{SYMsl2z} on the parameters
$\tau_0$ and $\beta$. In fact, after the passive action of
\eqref{SYMsl2z}, the resulting $\tau$ field does not have any easily
recognizable form. This breaks the $SL(2,Z)$ covariance of the Lunin-Maldacena
supergravity solution.

In the following section we will modify the Lunin-Maldacena solution in
such a way, that the passive action of the $SL(2,Z)$ transformation
\eqref{SYMsl2z} will be equivalent to the active $SL(2,Z)$ \eqref{strsl2z} on
the configuration itself. This reconciliation of the two $SL(2,Z)$
transformation on the string theory background would imply then that
there is a single $SL(2,Z)$ duality symmetry which acts on the gauge
theory side and on the string theory side. The action of this $SL(2,Z)$
on the two sides of the correspondence is in terms of the different
variables: $\tau_0$ and $\beta$ on the SYM side, and the supergravity
fields $\tau$ and the 2-forms on the string theory side. However the
symmetry is the same, and the $SL(2,Z)$ duality of the $\beta$-deformed
gauge theory implies the $SL(2,Z)$ duality of the string theory, and
vice versa.

\subsection{Periodicity properties}

The second observation concerns the fact that 
the  Lunin-Maldacena supergravity  dual \eqref{metrdef}-\eqref{f5def}
does not automatically exhibit 
the desired periodicity in the deformation parameter $\beta$
\be
\label{bdper}
\beta\, \to\,  \beta + 1 \ , \qquad \beta\, \to\,  \beta + \tau_0\ .
\ee
Here we need to make two clarifying comments following \cite{LM}: 
first, is that being a 
solution of supergravity, the configuration
\eqref{metrdef}-\eqref{f5def} is not even expected 
to be periodic under $\beta \to \beta+1$. This is because in going
from $\beta$ to $\beta+1$ one would have 
to go outside the region of $\beta\ll 1$ where we trust the
supergravity solution. The second comment is that 
in spite of this fact, the Lunin-Maldacena configuration for real
values of $\beta$ is formally invariant under 
a combination of the $\beta \to \beta+1$ and the $SL(2,Z)$ T-duality transformation.

Although with the help of the T-duality $SL(2,Z)$ of the
string theory, the Lunin-Maldacena
background is formally periodic under $\beta \to \beta +1$, 
this periodicity  is realized  
trivially\footnote{In particular, 
the weak-coupling instanton expression
for the $\tau$ field in \eqref{tauinsteff} is automatically periodic
and does not require 
any additional $SL(2,Z)$ transformation on the $S^5$ sphere generated
by instanton collective 
coordinates $\chi_{AB}$.} 
in the
SYM side. Hence,
one may wonder if there is a different way to implement the
periodicity condition $\beta \to \beta +1$ directly in the string dual background. 
In the following section we will extend the configuration in
\eqref{metrdef}-\eqref{f5def} to make it manifestly periodic under
both transformations in 
\eqref{bdper}.
The resulting $\beta$-periodic and $SL(2,Z)$ covariant configuration of
supergravity fields now faithfully represents the symmetries of the
$\beta$-deformed gauge theory. In the limit of $\beta \ll 1$ it will
also coincide with the Lunin-Maldacena supergravity dual. 

Our construction does not attempt to give a unique solution for the 
$\beta$-periodic and $SL(2,Z)$ covariant configuration. Our main point is to 
demonstrate that such configurations exist and to describe a
procedure for constructing 
such configurations.

We will
interpret the resulting configurations of supergravity fields as candidates
for the {\it string} dual background to the
$\beta$-deformed gauge theory. By construction our configurations are
$SL(2,Z)$ rather 
than $SL(2,R)$ covariant. We conjecture that 
the solution of equations arising from the string theory
effective action to
all orders in the $\alpha'$ expansion and in
string loops is covered by our general construction.

Finally, in section {\bf 4} we will show that our string configuration gives
predictions for the string theory effective action which are in
agreement with the multi-instanton calculations of appropriate
correlation functions in the $\beta$-deformed gauge theory. These
Yang-Mills calculations are taken from \cite{GK} and are carried out in
a weakly coupled gauge theory in the large-$N$ limit and to all orders
in the deformation parameter $\beta$.

Our conclusion will be presented in section {\bf 5}.

%%%%%%%%%%%%%%%%%%%%%%%%%%%%%%%%%%%%%%%%%%%%%%%%%%%%%%%%%%%%%%%%%%%%%%%%%%%%%%%%%
\section{String configuration, $SL(2,Z)$ covariance and periodicity in 
$\beta$}

In this section we will modify the supergravity dual of
Eqs.~\eqref{metrdef}-\eqref{f5def} to satisfy 
\begin{itemize}
\item the double periodicity
in $\beta$ of \eqref{bdper},  
\item the $SL(2,Z)$ covariance dictated
by Eq.~\eqref{strsl2z} when the parameters of the configuration are
transformed according to Eq.~\eqref{SYMsl2z}. Furthermore, we request that
\item the configuration reduces to the supergravity solution of
Lunin and Maldacena \eqref{metrdef}-\eqref{f5def} in the 
small $\beta$ limit $|\beta| \ll 1$ and
\item it  agrees with the Yang-Mills instanton prediction in the
limit of weak coupling $\tau_{02} \to \infty$ and large $N$. 
\end{itemize}

\subsection{Metric and form-fields}

The modification needed for the metric \eqref{metrdef}, two-forms
\eqref{B2def}-\eqref{C2def}, and the five-form fields \eqref{f5def} is
relatively straightforward. We have shown in the previous section
that the Lunin-Maldacena expressions for these fields already transform correctly 
under the $SL(2,Z)$. We need to keep this property and ensure that the fields
satisfy the required periodicity in $\beta$.
This 
can be achieved by replacing $\b$ in the original
expressions \eqref{GHdef}, \eqref{metrdef},
\eqref{B2def}-\eqref{f5def} 
by a function
$\cB(\b, \bb, \tau_0, \tb_0)$ 
\be
\label{B-tran}
\beta \, \longrightarrow \, \cB(\b,\bb,\tau_0,\tb_0) \ .
\ee
The function $\cB$ will be chosen to be 
doubly periodic in $\beta$, in agreement with \eqref{bdper},
and to transform under the $SL(2,Z)$ of \eqref{SYMsl2z} as
\be \label{B-transf}
\cB\left(\frac{\beta}{c \tau_0 +d}\, ,\, \frac{\bb}{c \tb_0 +d}\, ,\,
\frac{a \tau_0 +b}{c \tau_0 +d}\, , \frac{a \tb_0 +b}{c \tb_0 +d}
\right) \, =\, 
 \frac{\cB(\beta,\bb,\tau_0,\tb_0)}{c \tau_0 +d}\ .
\ee
We also require that 
\be \label{B-small-beta}
\cB \, \to \,  \b, \quad \mbox{for  $|\beta| \ll 1$ , $\tau_0$ arbitrary},
\ee
so that the Lunin-Maldacena solution is recovered in the small $\beta$ limit.
The above properties do not fix $\cB$ uniquely. To further constrain
$\cB$, we need a more detailed knowledge of the string
dynamics. 
As already mentioned earlier,
In this paper, we will be satisfied instead with
demonstrating that it is possible to construct such a solution.

The simplest solution can be constructed by 
assuming that
$\cB$ depends on $\beta$ and $\tau_0$ analytically, so that
\be \label{B-transfan}
\cB\left(\frac{\beta}{c \tau_0 +d}\,  ,\,
\frac{a \tau_0 +b}{c \tau_0 +d}
\right) \, =\, 
 \frac{\cB(\beta,\tau_0)}{c \tau_0 +d}\ .
\ee 
Thus,
in this case 
we will construct the string configuration using
elliptic functions. 
As we will show below, we can give a concise specification of
$\cB$ in this case.
If, 
on the other hand, the function $\cB$ is not analytic, 
then our approach outlined in \eqref{B-tran}-\eqref{B-small-beta}
still holds, we just cannot fully specify the form of $\cB$.
We will outline below how to construct 
the analytic function $\cB(\beta,\tau_0)$
with the desired properties.
The reader not interested in this discussion can skip directly to
 Eq.~\eqref{B-final}.
 
Elliptic functions
have the required double periodicity \eqref{bdper} 
and are meromorphic functions. 
For a review of elliptic functions one can consult
for example Ref.~\cite{akhiezer}.
It is known that the number of poles of an  elliptic function 
in a period parallelogram, counting multiplicity, cannot be less than
two. The simplest elliptic functions thus either have one pole of second
order, or two distinct poles of first order. Weierstrass function is of
the first type, while the Jacobi elliptic functions, sn, cn, dn, are of the
second type.  
In the following we will 
construct our 
candidate for the modified string background using the 
Weierstrass function.\footnote{Note 
that, unlike the Weierstrass function, the Jacobi elliptic functions 
do not have simple modular transformation like \eq{P-transf} below. }
We recall that the Weierstrass function is a function of two complex variables and
is defined by
\be
\label{Wedef}
\cP(\beta,\tau_0) := \frac{1}{\b^2} + \sum_{\l \in \Lambda/\{0\}}
\Big( \frac{1}{(\b-\l)^2} - \frac{1}{\l^2} \Big)\ ,
\ee
Here the lattice $\L = m + n \tau_0$ is generated by 1 and $\tau_0$.
The variable $\tau_0$ is defined on the upper half plane. 
By definition \eqref{Wedef}, the Weierstrass function satisfies 
the required periodicity in the variable $\beta$,
\be
\cP(\b+1,\tau_0)\, = \,\cP(\b+\tau_0,\tau_0) \,= \, \cP(\b,\tau_0)\ .
\ee
Furthermore, 
under the $SL(2,Z)$ transformation \eqref{SYMsl2z} 
of $\tau_0$ and $\beta$, 
$\cP$ transforms as a modular function of weight 2
\be \label{P-transf}
\cP\left(\frac{\beta}{c \tau_0 +d}\, ,\, 
\frac{a \tau_0 +b}{c \tau_0 +d}\right)\, =\,
(c\tau_0+d)^2 \, \cP(\beta,\tau_0)\ .
\ee
One can also define the derivative of the Weierstrass function 
with respect to the first argument.
The derivative is a modular function of weight 3,
that is
\be\label{P'-transf}
\cP'\left(\frac{\beta}{c \tau_0 +d}\, ,\, 
\frac{a \tau_0 +b}{c \tau_0 +d}\right) \, =\, 
(c\tau_0+d)^3 \, \cP'(\beta,\tau_0)\ .
\ee

For completeness, we note that
$\cP(\beta)$  satisfies the differential
equation
\be
\cP(\b)'{}^2 = 4 \cP^3(\b) - g_2 \cP(\b) -g_3\ ,
\ee
where the coefficients depend only on $\tau_0$,
\be
g_2 =\,  60 E_4 (\tau_0) \ , \quad g_3 =\,  140 E_6(\tau_0) \ . 
\ee
Here  $E_{2k}(\tau_0)$ denotes the Eisenstein series 
\be
E_{2k}(\tau_0)\, =\, \sum_{\l \in \L/\{0\}} \frac{1}{\l^{2k}}\ ,
\quad \mbox{for positive integer $k \geq 2$}\ .
\ee
$E_{2k}$ is a modular form\footnote{
A modular form is holomorphic on the upper half plane. Modular function
is needed to be meromorphic only.} 
of weight $2k$
\be
E_{2k}\left(\frac{a \tau_0 +b}{c \tau_0 +d}\right) \, =\,  
\left(c \t_0+d\right)^{2k} E_{2k}(\t_0)\ ,
\ee
and has the (weak-coupling) expansion near $q=e^{i\pi \t_0} \sim 0$,
\be
E_{2k} = 2\zeta(2k) + \frac{2 (2\pi i)^{2k}}{(2k-1)!}\sum_{m=1}^\infty
\s_{2k-1}(m) q^m\ ,
\ee
where  
$\zeta(s) =  \sum_{m=1}^\infty m^{-s}$ for ${\rm Re} s>1$ 
is the Riemann zeta function, and $\s_n(m) = \sum_{d|m} d^n$.

We can now construct the function $\cB$ in
Eq.~\eqref{B-transfan}
in terms of elliptic functions.
The general theory of elliptic function states that every elliptic function can
be  expressed in terms of the  $\cP$ and its
derivative $\cP'$  in the form \cite{akhiezer}
\be
f(\b) = R_1(\cP) + R_2(\cP) \cP',
\ee
where $R_1$ and $R_2$ are rational functions of their arguments. From the
transformation properties \eq{P-transf}, \eq{P'-transf} and 
\eq{B-transfan}, one can easily conclude that $B$ has to be of the form
\be \label{B-ansatz}
\cB \, =\,  \frac{a_1\,\cP'}{\cP^2 + E}\ ,
\ee
where $E=E(\tau_0)$ is a modular function of weight 4 
and $a_1=a_1(\tau_0)$ is a modular invariant function.
The formula above has to be consistent with the small $\b$ behaviour 
\eq{B-small-beta}.  
The asymptotic expansion of $\cP$ for small $\b$ is given by
\be
\cP(\b,\t_0) \, =\,  \beta^{-2} \, +\,  \frac{g_2}{20}\,\b^2 \,
+\, {\cal O}(\b^4) \quad \mbox{for $|\beta| \ll 1$ }.
\ee 
Thus \eq{B-small-beta} is reproduced if we set
\be
a_1 = -1/2\ .
\ee
As for $E$, it is natural to
require that 
it is analytic in $\tau_0$, so that $E$ is a modular form\footnote{In general 
$E(\tau_0)$ can also depend on the 
space-time coordinates, or at least on the coordinates 
$\bx$ 
of the deformed sphere.
For simplicity, we will assume that there is no such dependence, and $E(\tau_0)$
is a coordinate-independent modular form. In a sense, we
are building up the simplest generalization  of the Lunin-Maldacena 
solution which satisfies all the 
desired criteria. More general configurations are always possible.} of weight 4.
We note that
the set $\cM_{2k}$ of all modular forms of a given weight $2k$ 
is a finite dimensional
linear space over the complex field. The dimension of $M_{2k}$ is
given by 
\be
{\rm dim} \cM_{2k} = \bigg\{ 
\begin{array}{ll}
\left[k/6\right], &  \mbox{if $k \equiv 1$ (mod 6)} \\
\left[k/6\right] +1,  & \mbox{otherwise}\ .
\end{array}
\ee
In particular, $\cM_4 = \CC E_4$. Thus $\cB$ in Eq.~\eqref{B-tran} is
given by
\be \label{B-final}
\cB(\b,\t_0) \,=\, -\frac{1}{2}\, \frac{\cP'(\b,\t_0)}{\cP^2(\b,\t_0)\,+\, 
{\rm const} \cdot E_4(\t_0)}\ ,
\ee
where const is some constant.

It is always possible to construct more general solution $\cB$
without assuming analyticity. However as shown in \cite{GK} and
reviewed in section 4 below, the
instanton result \eq{tauinsteff} 
predicts that the string dilaton-axion field $\tau$ 
is analytic in $\beta$ (at least in the leading order at weak coupling).
It is less clear why the
the metric and the form-fields in the string dual background 
should be constructed using a  
holomorphic ``renormalization'' of $\beta$ to $\cB(\b,\t_0)$
as in our simple example above. In general, one can follow
the more general procedure in Eqs.~\eqref{B-tran}-\eqref{B-small-beta}
which involves non-analytic functions $\cB(\beta,\bb,\tau_0,\tb_0)$.
A better understanding of the string dynamics is then needed to fully
constrain the choice of $\cB(\beta,\bb,\tau_0,\tb_0)$.
 
In section {\bf 3.2} we will construct the dilaton-axion field $\tau$.
Before closing this 
subsection, we would like to make a few general comments.

First we want to comment on
the origin of the modifications of the supergravity dual.
We recall that the
Lunin-Maldacena solution with real
$\b$, can be obtained by the TsT transformation\footnote{Where T is a T-duality
and s is a shift.}
\cite{LM} from
the undeformed $AdS_5 \times S^5$ solution. 
Although $AdS_5\times S^5$ is an exact background of string theory, 
the generated solution does not admit the
required double periodicity\footnote{In particular, 
the dilaton-axion field $\tau$
in the Lunin-Maldacena background does not agree with the SYM instanton expression
\eqref{tauinsteff} unless $\beta$ is small.} 
and S-duality of the field theory.  
Hence, the supergravity dual cannot be an exact string theory background
even for real-valued deformations $\beta$. 
In fact, it is known that T-duality transformations for curved
backgrounds are generally modified by $\alpha'$ and string loop
effects. Therefore the modifications of the string solution from the original
Lunin-Maldacena solution can be attributed to the corrections to the form of the
T-duality transformation. We note that these corrections are becoming
more significant as 
$\b$ increases and this is the behaviour to be expected as large $\b$
corresponds to large curvature corrections ($\alpha'$).

It has been argued in \cite{BL,BJL} that for rational $\beta$, the
gauge theory is dual to an orbifold with a discrete torsion.\footnote{To
be precise, $Z_n\times Z_n$ orbifold when $\b= m/n$.} Furthermore,
Lunin and Maldacena have demonstrated \cite{LM} that their supergravity solution 
with rational values of the $\beta$ parameter is also related to the
orbifold description via an action of the $SL(2,Z)$ T-duality.
Our solution agrees with that of
Lunin-Maldacena for small $\beta$ and hence its connection with the
orbifold is guaranteed in this limit. However for generic rational
$\beta$, it is not clear how to relate our solution to the orbifold
description because the exact form of the T-duality 
rules is not available. 
In principle, the requirement of matching to the
orbifold description at rational values of $\beta$ 
should give and interesting constraint on the form of the T-duality
rule. It is possible that one may be able to derive the exact
T-duality rules in this way.

Finally, in spite of the fact that for rational values of $\beta$ the dual geometry
involves the orbifold, we expect that the geometrical description in terms of the 
$AdS_5\times \tilde{S}^5$ should hold throughout the parameter space, including these
rational values of $\beta$. In gauge theory in the conformal phase 
(which is what is relevant to our AdS/CFT correspondence) we do not expect to see
any discontinuities in the results as $\b$ becomes
rational.\footnote{All known instanton-generated 
as well as perturbative results in the SYM in the conformal phase
depend on $\beta$ smoothly. 
The story is different on the Coulomb branch where the F-term constraints
resemble the commutation
relation of a noncommutative torus and admit new solutions when $\b$
is rational.}

\subsection{Dilaton-axion pair}

Next we consider the  string dilaton-axion field $\tau = \chi + i e^{-\phi}$. 
In section {\bf 2.1} we found that the Lunin-Maldacena 
expression for $\tau$ in Eq.~\eqref{dilaxdef} did not transform covariantly 
under the $SL(2,Z)$. We want to construct the $\tau$ field which does.
Furthermore, the  resulting $\tau$ should
satisfy the weak-coupling and the small-$\beta$ limits. 
Writing $\tau$ as
\be \label{tau}
\tau = \tau_0 + \d\ .
\ee
The fact that
$\tau$ should agree with the original Lunin-Maldacena solution in the small
$\beta$ limit (for arbitrary $\tau_0$) gives:
\be
\d \, = \, \frac{i\beta^2}{2} \, g_{0E} \ , \qquad \quad\;\;
\mbox{for $|\beta| \ll 1$ and $\tau_0$ arbitrary}. \label{limit-1}
\ee
This equation is the leading order term in $\beta$ in the small $\beta$ expansion
of the Lunin-Maldacena $\tau$ field \eqref{dilaxdef}.
In addition to \eqref{limit-1} the modified configuration should
agree with the instanton result in the weak
coupling limit $\tau_{02} \to \infty$ (for arbitrary 
{\it complex} values of $\beta$):
\be
 \d \, =\, \frac{i g_{0E}}{2\pi^2}\, \sin^2\pi \b\ , \qquad \quad\;\;
\mbox{for $\t_{02} \to \infty$ and $\b$  arbitrary}.  
\label{limit-2}
\ee
Equation \eqref{limit-2} is the instanton prediction. We will discuss
its 
origin
in the following section. For now we only need to note that it is derived 
at weak-coupling  and in the large $N$ limit. What is 
important is that \eqref{limit-2} is valid for arbitrary
complex values of the deformation parameter $\beta$.

We note that $\tau$ is analytic in the parameters $\tau_0$ and $\b$ in these limits. 
We conjecture that the string field $\tau$ is
analytic in the parameters $\b$ and $\tau_0$ in general.
Therefore we will take
$\d = \d(\beta,\tau_0, \bx)$  a function of 
the parameters  $\beta$, $\tau_0$, as well as of the 
coordinates $\bx$ on the deformed sphere.
Writing 
\be
\d = \frac{u}{v}\ ,
\ee
it is easy to show that if $u,v$ transform under \eqref{SYMsl2z} as
\be \label{uv-transf}
u \to \frac{u}{(c\tau_0 +d)^2}\ , \quad v \to v + \frac{c u}{c\tau_0 +d}\ ,
\ee
then $\tau$ transforms as required by \eqref{strsl2z}
\be
\tau \to \frac{a \tau +b}{c \tau +d}\ .
\ee
Furthermore, $u$ and $v$ have to be constructed in such a way that the 
resulting $\tau$
satisfies the weak-coupling and the small-$\beta$ limits.

The transformation property of $v$ reminds us of that of the elliptic theta
function
\be
\th_1(z,\tau_0)\, := \, 2 q^{1/4} \sin\pi z \prod_{n=1}^{\infty}  (1-q^{2n})
\prod_{n=1}^{\infty} (1- 2 q^{2n} \cos 2 \pi z + q^{4n})\ , \quad q= e^{i
\pi \t_0}\ .
\ee
We recall that 
\be \label{theta-transf}
% \frac{\th_1(\frac{z}{c z+d}, \frac{a \t_0+b}{c \t_0 +d})}
% {\th_1'(0,\frac{a \t_0+b}{c \t_0 +d})} 
\frac{\th_1(z,\t_0)} {\th_1'(0,\t_0)}\, 
\rightarrow \, \frac{\exp(i \pi c z^2/(c \t_0+d)) }{c \t_0+d} \;
\frac{\th_1(z,\t_0)} {\th_1'(0,\t_0)}\ ,
\ee
which allows us to write a general solution to \eq{uv-transf} as follows:
\be \label{uv-ansatz}
u \,=\, z^2\ , \qquad 
v\, =\, h +\frac{1}{i \pi} \ln\Big(\frac{\th_1(z,\t_0)} 
{z\;\th_1'(0,\t_0) } \Big) \ .
\ee
Here $z=z(\b,\tau_0,\bx)$ is required to transform as\footnote{We have
chosen to define $u$ in terms of $z$, rather than use $\sqrt{u}$ in the
argument of  
the $\theta$-function
in the second equation in
\eqref{uv-ansatz}. This is because the function $\sqrt{u}$ is
multi-valued. A branch for the square root has to be chosen in order to
define $v$ in \eq{uv-ansatz}. However, the branch cut is not invariant
under \eq{z-transf}. Therefore we conclude that 
\eq{uv-ansatz} is well-defined only if $u$ is a
complete square and hence $z$ is single-valued.}
\be \label{z-transf}
z \, \to \, \frac{z}{c\t_0+d}\ ,
\ee
and $h(\b,\tau_0,\bx)$ is a modular invariant function. 
We also choose the branch of the logarithm where $\ln 1 =0$. 

As in the above,
one may represent the elliptic function $u$ as 
a rational function of $\cP$ and $\cP'$. 
However only a definite combination of them can be written as a
complete square $z^2$ with  the required weight. This gives a form
similar to \eq{B-final} for $\cB$. It is natural to assume that 
$z$ is simply proportional to
$\cB$  
\be
z\, =\,  b_1 \,\cB\ .
\ee
where $b_1(\t_0)$
is a modular invariant function.\footnote{Again, on general grounds, 
it is possible that $b_1(\t_0)$ also depends on the space-time 
coordinates $\bx$. In what follows we will assume 
a simple scenario where $b_1$ depends only on $\t_0$.} 
$z$ is doubly periodic in $\b$.

Now we examine the limits. In the leading order at small $\beta$, we find
\be
z = \b\, b_1 \ ,   \quad 
v= h(0,\t_0,\bx)\ . 
\ee
Thus the matching with the Lunin-Maldacena expression for $\tau$ in 
the small $\b$ limit gives 
\be \label{LM-cond}
h(0,\t_0,\bx) \, =\,  -2i b_1(\t_0)^2 /g_{0E}\ .
\ee
Next we look at
the weak coupling limit. We have 
\be
\cP(\b,\t_0) \, =\,  \pi^2( \frac{1}{\sin^2 \pi \b} - \frac{1}{3})\ ,
\qquad  E_4(\tau_0) \,=\,  \frac{\pi^4}{45}\ , \qquad
\tau_{02} \to  \infty\ ,
\ee 
and thus $\cB$ is independent of $\t_0$ in the limit. 
Therefore if we choose
$b_1$ such that it becomes infinite in the weak-coupling limit, and also set
$h$ in \eqref{uv-ansatz} to be
\be \label{h-f}
h \, =\, z^2 f
\ee 
for some modular function $f$ of weight 2  
which remains finite in the limit, 
then the $\log$ term in 
$v$ in \eqref{uv-ansatz}
can be neglected. Hence 
$u/v = 1/f$
in this limit. 
This can be achieved by taking 
$b_1$ to be, for example, the modular  function
$J(\t_0):= g_2^2/(g_2^3 -27g_3^2)$
since it has a double pole at $q =0$
\be
J(\t_0)\, = \, (1728)^{-1}q^{-2} + {\rm finite} \ .
\ee

To match with the
instanton result \eq{limit-2}, the function $f$ has to satisfy
\be
\lim_{\tau_{02} \to \infty} f(\b,\t_0,\bx) 
\, =\, \frac{-2i \pi^2}{g_{0E} \sin^2\pi \b} \ .
\ee
This can be satisfied by
\be
f(\b,\t_0,\bx) \, =\,   \frac{-2i}{g_{0E}} (\cP + G(\tau_0))\ ,
\ee
where $G(\tau_0)$ is a modular function of weight 2 such that
$\lim_{\tau_{02} \to \infty} G(\t_0) = \pi^2/3$. This can be constructed
from the Eisenstein series, for example, 
$G (\t_0) \,=\,  \frac{7 E_{6}}{2 E_{4}}$.
One can check easily that \eq{LM-cond} is also satisfied.

Summarizing, we have constrained the modified string configuration using 
the requirement of double periodicity in $\beta$, the $SL(2,Z)$
symmetry, and matching to the known asymptotic form 
of the supergravity solution in the small $\beta$
and in the weak coupling limits. 
The result is not uniquely fixed
without further input of the full
string dynamics. A particularly simple solution can be
written down with the assumptions of analyticity. The result is 
that for the metric and the 2-form fields, one has to
replace the $\beta$ parameter in the original Lunin-Maldacena configuration 
by the function $\cB$
of \eq{B-final}
\be 
\cB(\b,\t_0) \, =\, 
-\frac{1}{2}\,\frac{\cP'(\b,\t_0)}{\cP^2(\b,\t_0)\,+\,  {\rm const}
  \cdot E_4(\t_0)} \ . 
\ee
For the dilaton-axion field we have the following expression,
\be
\tau(\b,\t_0,\bx)\, = \, \tau_0 \, -\, i \Bigg[
\frac{- 2}{g_{0E}(\bx)} 
\bigg(\cP(\b,\t_0) + G(\tau_0)\bigg) 
+\frac{1}{ \pi z^2} \ln\bigg(\frac{\th_1(z,\t_0)} {z\;\th_1'(0,\t_0) }
\bigg)\Bigg]^{-1},
\ee
where
\be
z\, = \, b_1(\t_0)\,  \cB(\b,\t_0)\ .
\ee
Here 
 $b_1$ is a modular invariant function which
blows up in the weak coupling limit $\t_{02} \to \infty$. A simple choice for
$b_1$ would be $b_1= J(\t_0)$.
Finally, $G(\t_0)$ is a modular function of weight 2.

We remark that in establishing the classical integrability
of the Lunin-Maldacena background (for real $\beta$) in \cite{integr2}, 
the knowledge of the
dilaton-axion pair was never used, and apart from the dilaton-axion pair, our
proposed string background is the same as Lunin-Maldacena with the substitution
$\beta \to \cB$. Therefore the classical
integrability (for real $\beta$) apply equally 
for any $\cB(\b,\bb,\t_0,\tb_0)$ with the property that $\cB$ is
real whenever $\b$ is real.

%%%%%%%%%%%%%%%%%%%%%%%%%%%%%%%%%%%%%%%%%%%%%%%%%%%%%%%%%%%%%%%%%%%%%%%%%%%%%%%%%
\section{String theory effective action and instantons}

The string effective action $S_{\rm IIB}$ is related via the
AdS/CFT holographic formula \cite{GKP,Witten150} to
correlation functions in the gauge theory,
\begin{equation}
 \exp-S_{\rm IIB}\left[\Phi_{\cal O};J\right]\, =\, 
 \left\langle \exp\,\int\, d^{4}x \, J(x){\cal O}(x)\,
\right \rangle \ .
\label{corr}
\end{equation}  
Here $\Phi_{\cal O}$ are Kaluza-Klein modes of the supergravity fields 
which are dual to composite gauge theory operators $\cal O$.
The boundary conditions of the supergravity fields 
are set 
by the gauge theory sources on the boundary
of $AdS_5$.

The effective action $S_{\rm IIB}$ of type IIB string theory is
invariant under the $SL(2,Z)$ transformations \eqref{strsl2z}. The
action of this symmetry leaves the metric invariant, but acts upon the
dilaton-axion field $\tau$, as well as on the 2-form fields. It is
well-known that in the supergravity approximation, the action is
invariant under the full $SL(2,R)$ symmetry, however the
higher-derivative corrections to it, as well as the string-loop
corrections break the $SL(2,R)$ symmetry down to $SL(2,Z)$.

At leading order beyond the
Einstein-Hilbert term in the derivative expansion, the IIB effective
action is expected to contain \cite{GG1}, \cite{BG}
the ${\cal R}^4$ term 
\be
(\alpha')^{-1} \int d^{10}x \,\sqrt{-g_{10}} \,e^{-\phi/2}
\, f_4(\tau,\bar\tau) \,
 {\cal R}^4 \ ,
\label{twoterms}
\ee
as well as its many superpartners, including
a totally antisymmetric 16-dilatino
effective vertex of the form 
\begin{equation}
(\alpha')^{-1}\int d^{10}x\,\sqrt{-g_{10}} \,e^{-\phi/2}\,
f_{16}(\tau,\bar\tau)\,\Lambda^{16}\ + \ \hbox{H.c.}
\label{effvert}\end{equation}
D-instanton contributions in supergravity contribute precisely to
Eqs.~\eqref{twoterms}-\eqref{effvert} and their superpartners, i.e. to
the leading order higher-derivative corrections to the classical IIB
supergravity \cite{BG}. The D-instanton contribution to an $n$-point
interaction of supergravity fields comes from a tree level Feynman
diagram with one vertex located at a point $(x_0,\rho,\hat\Omega)$ in
the bulk of $AdS_{5}\times \tilde{S}^{5}$. The diagram also has $n$
external legs connecting the vertex to operator insertions on the
boundary. We will outline below how to single out these D-instanton
contributions in \eqref{twoterms}-\eqref{effvert}.

We further note that the higher-derivative corrections 
Eqs.~\eqref{twoterms}-\eqref{effvert} must respect the $SL(2,Z)$ 
of Eq.~\eqref{strsl2z}.
Under these transformations supergravity field components $\Phi$ acquire 
(discrete) phases,
\be
\Phi\, \longrightarrow \, 
\left(\frac{c\tau+d}{c\bar\tau+d}\right)^{-\,q_{\Phi}/2} \Phi \ ,
\label{phases}
\ee
The charge $q_{\Phi}$ for the dilatino is $3/2$ and for the ${\cal R}$ 
field it is zero.

Equations \eqref{twoterms}-\eqref{effvert}
are written in the string frame with the coefficients
$f_n(\tau,\bar\tau)$ being the modular forms of weights $((n-4),-(n-4))$
under the $SL(2,Z)$ transformations \eqref{strsl2z},
\be
f_n(\tau,\bar\tau) \, :=\,
f^{\,(n-4), -(n-4)} (\tau,\bar\tau) \, \longrightarrow \, 
\left(\frac{c\tau+d}{c\bar\tau+d}\right)^{n-4}
f^{\,(n-4), -(n-4)} (\tau,\bar\tau)\, .
\ee
The modular properties of $f_n$ precisely cancel the phases of fields in
\eqref{phases} acquired under the $SL(2,Z).$ Thus the full string
effective action is invariant under the $SL(2,Z)$ and this modular
symmetry ensures the S-duality of the type IIB superstring.

The modular forms $f_n$ have been constructed by Green and Gutperle 
in \cite{GG1}.
In the weak coupling expansion the expressions for $f_n$
contain an infinite sum of exponential terms
\begin{equation}
e^{-\phi/2}\, f_n \, \ni \,
\sum_{k=1}^{\infty}\, {\rm const}\cdot\left({k\over G^{1/2}\,g^2}\right)^{n-7/2}\, 
e^{2\pi ik\tau}\, \sum_{d|k}{1\over d^2}\ ,\label{foksh}
\end{equation}
It is clear that in the sum above each term corresponds to a contribution of a
D-instanton of charge $k$. On the other hand, $k$-instanton contributions can
also be calculated directly in gauge theory.
It was shown recently in Ref.~\cite{GK} that
each of the terms in the sum in the expression above
can be identified
with a contribution of an instanton of charge $k$ in the
$\beta$-deformed SYM theory.\footnote{The result of \cite{GK} is a
generalization to the $\beta$-deformed case of earlier instanton
calculations carried out in the undeformed $\N=4$ SYM. These
calculations have been performed in \cite{BGKR,DKMV} at the 1-instanton
level and in \cite{DHKMV,MO3} for the general $k$-instanton case.
Remarkably, these gauge theory results, including the results of
\cite{GK} in the $\beta$-deformed theory are in precise agreement with
the supergravity predictions for the effective action
Eqs.~\eqref{twoterms}-\eqref{foksh}. In what follows it will be
sufficient to concentrate only on reproducing the exponential factors
$e^{2\pi ik\tau}$ on the right hand side of Eq.~\eqref{foksh}.} This
precise identification was performed in \cite{GK} at weak coupling in
the large-number of colours $N \gg 1$ and the small-deformation $\beta
\ll 1$ limits. The latter limit in particular, was necessary to ensure
that the Lunin-Maldacena supergravity dual remains a valid approximation
to string theory \cite{LM}. Here we will generalize these results to
arbitrary values of the deformation parameter $\beta$.

We now consider the exponential factors $e^{2\pi ik\tau}$ in the
Eq.~\eqref{foksh}, and compare them with the relevant (generic) part of
the multi-instanton contributions to correlators of the $\beta$-deformed
gauge theory calculated in \cite{GK}. 
The generic Yang-Mills $k$-instanton contributions have been recently calculated in  \cite{GK}
and are of the form:
\be
{\cal F}_k := \, e^{-k{8\pi^2\over g^2}+ik\theta}\,
\left(1\, -\, 4\, \sin^2 (\pi \beta)\, Q\right)^{k(N-2)}\ .
\label{Fdef1}
\ee
This expression was derived \cite{GK} in the weak-coupling limit in gauge theory 
and is valid for arbitrary values of the complex deformation parameter $\beta$.
Here $Q$ is the same
function of the $\mu_i$ coordinates on the deformed sphere $\tilde{S}^5$
as the one appearing in \eqref{Qdef}. These coordinates and the sphere
$\tilde{S}^5$ itself arise in the Yang-Mills instanton approach from the
bosonic collective coordinates $\chi_{AB}$ of the instanton, which are
used to bi-linearize the term in the instanton effective action of the
fourth order in fermionic collective coordinates. We refer the reader to
sections 4 and 8 of Ref.~\cite{GK} (or section 4 of Ref.~\cite{MO3}) for more
detail.

This $k$-instanton factor ${\cal F}_k$ is supposed to match with the 
$k$ D-instanton term  $e^{2\pi ik\tau}$ in \eqref{foksh}. We thus have, 
\be
\label{tauinstp}
\tau \, =\, \tau_0 \, +\, \frac{N}{2\pi i}\, \log\,
\left(1-\,4\,Q\,\sin^2 \pi \beta\right)\ . 
\ee
We can now simplify the instanton prediction above if we recall that the
instanton measure includes the integration over all instanton collective
coordinates, including the integration over the sphere $\tilde{S}^5$.
These integrations are supposed to be carried out in the limit of large
number of colours, $N \to \infty$. The function $Q$ on the right hand
side of \eqref{tauinstp} is a function of the collective coordinates,
which themselves are integration variables. Schematically, we have
\bea
\int\,  d\mu_1 \,d\mu_2\, d\mu_3 \, &&\delta(\mu_1^2+\mu_2^2+\mu_3^2)
\, e^{2\pi ik\tau} \\ 
 =\,
&&\int\,  d\mu_1 \,d\mu_2\, d\mu_3 \, \delta(\mu_1^2+\mu_2^2+\mu_3^2) \,
\exp\left[2\pi i k\tau_0\, +\, kN\,\log\, \left(1-\,4\,Q\,\sin^2 \pi
    \beta\right) \right]\ . \nonumber
\eea
In the $N \to \infty$ limit, the integral can be performed in the
saddle point approximation. This selects the dominant 
value of the function $Q(\mu_i)$ to be $\langle Q \rangle \sim 1/N \ll 1$.
This implies that the logarithm in \eqref{tauinstp} can be
power-expanded and only the leading order term in $Q$ should be kept 
in anticipation of the integrations over collective coordinates.
This gives the effective instanton prediction for 
$\tau$ field in string theory in the form:
\be
\label{tauinsteff}
\tau \, =\, \tau_0 \, +\, i\, \frac{g_{0E}}{2\pi^2}\, \sin^2 \pi \beta \ ,
\ee
where we have used $g_{0E} =\, 4 \pi N Q $
as in \eqref{Qdef}.
As expected, the dependence on $\beta$ is periodic as $\beta \to \beta+1$.
The second type of $\beta$-periodicity, $\beta \to \beta+\tau_0$,
clearly  cannot be seen in this semiclassical instanton result. 
Indeed, at week coupling $\tau_{02} \to \infty$, and
the second periodicity is lost.

Equation \eqref{tauinsteff} is the instanton prediction which we have used in
\eqref{limit-2} in constraining the form of the string configuration in
the previous section.

%%%%%%%%%%%%%%%%%%%%%%%%%%%%%%%%%%%%%%%%%%%%
\section{Conclusions}

In this paper we have constructed the generalization of the
Lunin-Maldacena supergravity solution dual to the $\beta$-deformed
conformal gauge theory. Our modified configuration satisfies the
criteria outlined in the beginning of Section 3. In particular, this
configuration has a double periodicity in the deformation parameter
$\beta$. It also transforms covariantly under the $SL(2,Z)$ duality of
string theory when the parameters $\tau_0$ and $\beta$ are transformed
under the gauge-theory $SL(2,Z)$. This reconciles the $SL(2,Z)$
Montonen-Olive duality of the 
$\beta$-deformed SYM with the string theory
$SL(2,Z)$ invariance.

In the $\beta$-deformed case, the supergravity background receives corrections
in string theory and thus should not identified with the exact string background.
Since our configuration is fully consistent with the symmetries of the theory
and since it
transforms under the $SL(2,Z)$ rather than the
$SL(2,R)$ symmetry, we expect it to represent the string theory (and not
the supergravity) background. We propose that in comparing to the SYM side
one should use this string background configuration in the string theory effective action.

We have encountered a certain degree of
freedom in our construction. This freedom cannot be fixed from the
symmetry requirements alone, nor by the matching with the known limits
in supergravity and gauge theory. We have presented the simplest form of
the solution. More general solutions can be easily constructed, and they
should be tested when better knowledge of the full dual string theory
dynamics is available.

\bigskip
\bigskip

\centerline{\bf Acknowledgements}

We acknowledge useful discussions with
Adi Armoni, Ed Corrigan, Tim Hollowood, George Georgiou,
Asad Naqvi and Carlos N{\'u}{\~n}ez at the outset of this work.
We also thank Oleg Lunin and Juan Maldacena for comments on this paper.
The research of CSC
is supported by EPSRC through an Advanced Fellowship. VVK is supported by a PPARC
Senior Fellowship.

\bigskip

%\newpage
%%%%%%%%%%%%%%%%%%%%%%%%%%%%%%%%%%%%%%%%%%%%%%%%%%%%%%%%%%%%%%%%%%%%%%%%%%%%%%%%%
%\startappendix

%\Appendix{}

\newpage
%%%%%%%%%%%%%%%%%%%%%%%%%%%%%%%%%%%%%%%%%%%%%%%%%%%%%%%%%%%%%%%%%%%%%%%%%%%%%%%%%%

\end{document}